# Photochemistry in hot $H_2$-dominated exoplanet atmospheres

Benjamin Fleury, Murthy S. Gudipati[*], Bryana L. Henderson, Mark Swain

Science Division, Jet Propulsion Laboratory, California Institute of Technology, 4800 Oak Grove Drive, Pasadena, California 91109, USA

[*]Corresponding author: murthy.gudipati@jpl.nasa.gov

Pages: 32

Tables: 3

Figures: 10

**Abstract**

Photochemistry has the potential to substantially impact the atmospheric composition of exoplanets with consequences on the radiative transfer, thermal structure and dynamics of the atmospheres, particularly in UV-rich stellar environments. Here, we present the results of a first laboratory experimental simulation of photochemistry in carbon-rich exoplanet atmospheres at elevated temperatures. Evolution of gas-phase molecular composition was quantitatively monitored with infrared spectroscopy and mass spectrometry. We found that $H_2$/CO gas compositions can change significantly from thermal equilibria compositions when irradiated with Ly-α photons at temperatures ranging from 600 K to 1500 K. Carbon dioxide and water were found to be the main products caused by photolysis, while formation of methane was also observed to a lesser extent. We find that photochemistry efficiency is strongly correlated with increasing temperature. Our finding that water is efficiently produced by photochemistry in a super Solar C/O=1 environment, representing C enhancement relative to solar values C/O ratio = 0.54, has significant implications for the interpretation of many exoplanet transmission spectra. We also find the formation of an organic solid condensate at 1500 K and under Ly-α UV-radiation, confirming the possibility of forming photochemical hazes in hot-Jupiter exoplanet atmospheres with an enhanced C/O ratio compared to Solar.

# 1. Introduction

A part of known exoplanets are hot (800 – 3000 K) giant planets with short orbital periods and correspondingly small semi-major axis values (Moses 2014; Madhusudhan et al. 2016), which are exposed to high temperatures as well as high UV fluxes. As indicated by a wide range of bulk densities, their atmospheres are likely diverse in composition, and are influenced by several factors: the temperature profile, the elemental C/O ratio, the metallicity, the host star type, and UV flux (Moses et al. 2013a; Moses et al. 2013b; Miguel & Kaltenegger 2014; Venot et al. 2014). The gas giant planets, ranging in extremes from hot-Jupiters to the cooler sub-Neptunes, are expected to have atmospheres mainly composed of dihydrogen ($H_2$) and Helium (He) (Madhusudhan, et al. 2016). Observations during transit and eclipses have revealed the presence of carbon, oxygen and nitrogen-bearing molecules such as carbon monoxide (CO), carbon dioxide ($CO_2$), methane ($CH_4$), ammonia ($NH_3$), and water ($H_2O$) (Tinetti et al. 2007; Swain et al. 2009a; Swain et al. 2009b; Swain et al. 2010; de Kok et al. 2013; Wakeford et al. 2013; MacDonald & Nikku 2017). Extensive H escape (Vidal-Madjar et al. 2003; Ehrenreich et al. 2015), presumably reflecting the underlying $H_2$-dominated nature of the atmospheres, has been observed as well.

For exoplanets with T < ~1500 K, disequilibrium chemistry such as transport-induced quenching and photochemistry can affect seriously the atmospheric gas-phase composition (Moses et al. 2011; Venot et al. 2012; Moses 2014) and then impact the radiative properties, thermal structure and dynamics of the atmospheres (Moses 2014; Drummond et al. 2016). Further, some observations suggest that aerosols are present in the atmospheres of exoplanets such as the super-Earth GJ 1214b (Kreidberg et al. 2014), the hot-Neptune GJ 436b (Knutson et al. 2014) and several hot-Jupiters (Demory et al. 2013; Iyer et al. 2016; Sing et al. 2016). Aerosols can affect the albedo, the thermal structure and the chemistry of the exoplanet atmospheres similar to what is observed in our solar system (Marley & McKay 1999; West et al. 2014). However, whether the observed aerosols are mainly condensate clouds (Parmentier et al. 2016) or photochemical hazes (Marley et al. 2013) remains largely unknown. It is also not clear what percentage of these aerosols could be organic (carbon-based) or inorganic (silicates, etc.). We focus here on organic aerosol formation.

Several chemical and microphysical models have been developed over recent years to study disequilibria processes such as quenching and photochemistry (Zahnle et al. 2009; Line et al. 2010; Line et al. 2011; Moses, et al. 2011; Venot, et al. 2012; Venot et al. 2015) and aerosol formation in exoplanet atmospheres (Zahnle et al. 2016; Gao et al. 2017; Lavvas & Koskinen 2017; Kawashima & Masahiro 2018). However, corresponding laboratory data for these processes are sparse and laboratory investigation of the photochemistry and aerosol formation in exoplanet atmospheres is crucial for their characterization and the interpretation of existing and future observations (Fortney et al. 2016). Only a few laboratory experiments thus far have focused on aerosol formation and aerosol properties in Earth-like exoplanets (Gavilan et al. 2017) or super-Earth and mini-Neptune-like exoplanets (He et al. 2018; Hörst et al. 2018). These studies highlight the efficient formation of photochemical hazes in a large diversity of atmospheric conditions with different physical and chemical properties. However, these experiments were limited to planets with relatively cool atmospheres (T < 600 K) and do not mimic the higher temperature exoplanet atmospheres such as in hot-Jupiters, which represent a majority of currently observed exoplanet atmospheres. In addition, these experiments used plasma discharges as energy sources to simulate the chemistry occurring in exoplanet atmospheres. Although plasma discharges are more efficient at generating aerosol analogs in the laboratory, chemistry in planetary atmospheres is mainly driven by stellar UV radiation. Hence, our work focuses on aerosol formation at elevated temperatures (600 K – 1500 K) and under Ly-α UV-radiation.

In our study, we have chosen to investigate UV photochemistry and thermal equilibrium chemistry involving $H_2$ and CO, the two most abundant molecules in hot-Jupiter type exoplanet atmospheres (excluding He, which is chemically inert). The combination of $H_2$ and CO represents the simplest plausible hot Jupiter atmosphere with T > 1000 K and corresponds to a C/O ratio of 1 representing C enhancement relative to solar values. These atmospheres are the focus of our study. The existence of this type of exoplanet has been suggested after the observation of the hot-Jupiter WASP-12b (Madhusudhan et al. 2011), even though the C/O ratio itself is still debated (Madhusudhan 2012; Mandell et al. 2013; Moses, et al. 2013b; Swain et al. 2013). However, recent work by Brewer et al. (2017) suggests the enhancement of C/O in hot Jupiter atmospheres may be common. Thermal chemical models predict that these exoplanets have $H_2$-dominated atmospheres with CO as the main carbonaceous species for atmospheres with temperatures higher than 1000 K (Moses, et al. 2013b; Venot, et al. 2015). These atmospheres are expected to contain heavier and more complex C-bearing molecules (Venot, et al. 2015), making them an important target for the study of the formation of photochemical hazes. Finally, previous works have demonstrated that the ultraviolet absorption cross-section of molecules is temperature dependent (Venot et al. 2013; Venot et al. 2018), and has a large impact on the photodissociation efficiency and the photochemistry (Venot, et al.

2015; Venot, et al. 2018) in planetary atmospheres. For this reason, we investigate the effect of temperature on photochemical evolution of exoplanet atmospheres.

## 2. Experimental setup and analytical protocols

### 2.1. Cell for Atmospheric and Aerosol Photochemistry Simulations of Exoplanets (CAAPSE)

Figure 1 presents the scheme of the Cell for Atmospheric and Aerosol Photochemistry Simulations of Exoplanets (CAAPSE) experimental setup. This is a new instrument, custom-built at the Jet Propulsion Laboratory to simulate photochemistry and formation of aerosols in exoplanet atmospheres. The reaction cell is 48 cm long and composed of a 38 cm long quartz or alumina tube with an internal diameter of 6.1 cm, and is closed with two $MgF_2$ windows mounted on stainless-steel flanges with high-temperature O-ring seals. Alumina has a better resistance to higher temperatures than quartz and can be used to simulate exoplanet atmospheres with a higher temperature. However, alumina is more porous than quartz, enhancing the surface area of gas-solid contact significantly and possibly also catalytic effects. To evaluate this effect, we performed comparative experiments using the two types of tubes. A customized STT-1600C (SentroTech) oven is used to heat the cell up to 1773 K (1500 °C). The maximum allowed temperature is 1600 °C on this instrument, however we opted to stay well below that threshold to avoid unwarranted processes such as melting of the O-rings etc. $MgF_2$ windows are the only optical material that is transparent both for Ly-α photons and for infrared spectral wavelengths. However, $MgF_2$ is thermally rated only up to 300 °C. For this reason, these windows are water-cooled to avoid any thermal damage while allowing the rest of the cell to operate at a higher temperature. Before each experiment, the cell is heated at 1173 K and pumped down to $3\times10^{-7}$ mbar for 24 hours to degas it and remove adsorbed water or other impurities.

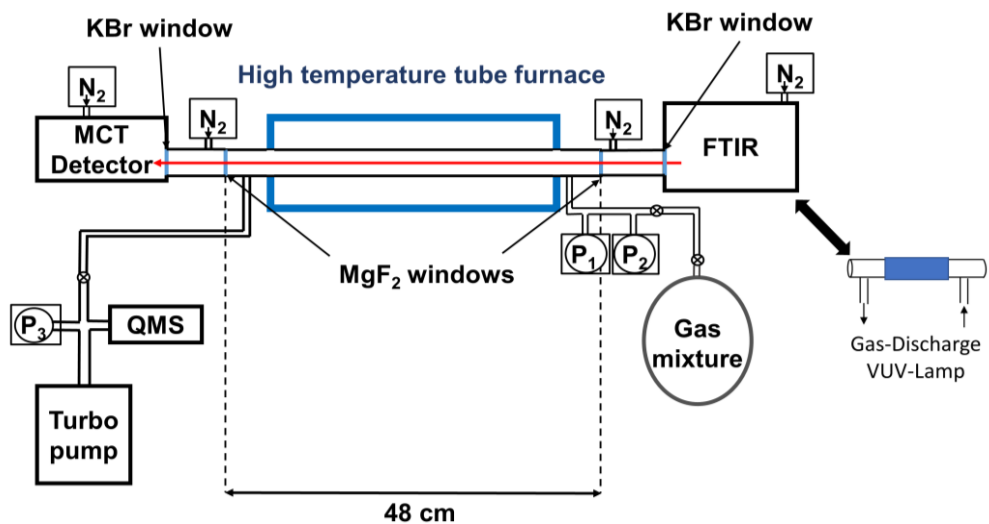

**Figure 1: Scheme of the CAAPSE experimental setup at JPL.**

Our mixtures have been prepared based on atmospheric composition calculated using a thermo-chemical equilibrium model that has a C/O ratio of 1.1 and a temperature of 1000 K in Venot et al. (2015), making the CO molecule a good choice to represent this C/O ratio. We chose to use only the two most abundant species calculated in the model of Venot et al., 2015, $H_2$ and CO, to limit the experimental complexity, which results in the exclusion of nitrogen-bearing molecules such as $N_2$ as well as He, which is chemically inert. Although the $N_2$ mixing ratio is very low in Hot-Jupiter atmospheres, further studies would be necessary to determine if nitrogen chemistry may have an effect on aerosol formation and properties, as it does on Titan, for example. To prepare the initial gas mixture, we filled $H_2$ (Airgas, 99.9999%) and $^{13}CO$ (Cambridge Isotope Laboratories, 99.9%) into a glass bulb of 2 L volume previously pumped down to $1\times10^{-3}$ mbar on a glass gas-mixing vacuum line. Typical mixtures obtained were $H_2$ (99.7 %) and $^{13}CO$ with a mixing ratio of $3\times10^{-3}$ (0.3 %) by volume. Total pressure in the glass bulb was kept under 600 mbar and the gas mixture was kept at room temperature for at least 12 hours before loading into the CAAPSE cell. Isotopically-labeled carbon monoxide ($^{13}CO$) was used to isolate the effect of any contamination (spectral contamination due to atmospheric gases, atmospheric gas leaks into the cell, outgassing of the inner surfaces, etc., discussed in Section 3.2.3) from the thermal and photochemical processing within the cell on the initial gas-mixture. The quartz and alumina tubes were used only with gas mixtures made of $H_2$ and $^{13}CO$ in order to avoid possible cross-contamination. The reaction cell was filled with 15 mbar of the initial gaseous mixture and heated at 5 K·min$^{-1}$ to different oven temperatures, starting from room temperature of 295 K and subsequently at 573 K, 873 K, 1173 K, 1273 K, 1373 K, and 1473 K. Although the gas mixture composition we used is only representative for Hot-Jupiter like atmospheres with a T > 1000 K and a C/O ratio of 1, we performed control experiments at 573 K and 873 K to obtain more information on how the gas temperature affects the chemistry for a given atmospheric composition. Such data are indeed crucial to understand chemical processes in our experimental setup and to better understand the effect of the gas temperature on the thermochemistry and photochemistry. The temperature of the tube was monitored using three type-B thermocouples distributed across the furnace at a spacing of 4.5 cm and read into a computer with MadgeTech 4 software. During heating, evolution of the gas mixture composition was monitored continuously by infrared (FTIR) spectroscopy.

After attaining the desired temperature, the gas mixture was kept at that temperature until the changes seen in the infrared spectra were minimal (see Section 3.1). From that point on, the heated gaseous mixture was irradiated using a microwave discharge lamp with a 1.2 mbar continuous flow of $H_2$ (Airgas, 99.9999%) powered by a microwave generator (OPTHOS) set to 70 W. The lamp was separated from the reaction chamber by a $MgF_2$ window that cuts off wavelengths shorter than 110 nm. Under these conditions, the lamp produces a continuous emission of photons in the UV at 121.6 nm (Ly-α) and a broad but weaker emission in the 140–170 nm range. It mimics exoplanet atmospheres irradiated by UV-photons from their UV-rich stars, with a predominance of Ly-α (121.6 nm) in the vacuum ultraviolet (VUV, λ < 200 nm), especially for planets orbiting around M stars (France et al. 2013; Miguel et al. 2015). The photon flux of the lamp is estimated to be ~$10^{13}$ photons·cm$^{-2}$·s$^{-1}$ at 121.6 nm (Ligterink et al. 2015). Although these photons are not energetic enough to directly dissociate or ionize $H_2$ and CO, previous studies have demonstrated that photochemistry can be driven through photoexcitation of CO (Liuti et al. 1966b, 1966a; Dunn et al. 1973; Vikis 1982; Roussel & Back 1990). As the CAAPSE cell has only two optical windows, we needed to exchange the FTIR with the microwave discharge lamp for the irradiation. Thus, it was not possible to monitor continuously the evolution of the gas mixture by IR spectroscopy during the irradiation but only at the end of the irradiation, when the lamp was removed.

The pressure in the CAAPSE cell was monitored at each step of the experiments by a CDG-500 capacitance diaphragm gauge (Agilent). The pressures measured in the cell and the total time since the gas mixture loading into the cell are summarized in Table 1 at the different steps of the experiments for each set temperature. The thermal expansion of the gases during the heating of the cell involved an increase of the pressure inside the cell and as seen from the Table 1, the gas mixture does not follow a strict ideal gas proportionality ($P_2/P_1 = T_2/T_1$) at constant volume. This is due to the existence of a temperature gradient along the tube (see Figure 2), which also led to a pressure gradient in the cell. At elevated temperatures beyond 600 K we noticed an increase of the pressure of ~1 mbar, after the irradiation for 18 hours, which could be due to a leak (see Section 3.2.3 for further discussion). It should be noted that once the gas mixture at a total pressure of 15 mbar is loaded into the reaction cell, all the valves are closed and the cell is subjected to atmospheric pressure at every potential point of leak (O-rings). Hence it is very difficult to avoid leaks in such a system subjected to extreme temperatures. We took all the precautions to minimize these leaks. As will be seen in the following, small leaks do not appear to make any significant impact on our results, and the choice of using isotopically labeled $^{13}CO$ enables us to trace such leaks and quantify them.

**Table 1: Summary of the total time since loading the $H_2$:$^{13}CO$ (99.7%:0.3%) gas-mixture into the cell at room temperature vs. pressure in the CAAPSE cell at the different steps of the experiments for each set temperature.**

|  | Initial condition at 295 K | | Condition at the thermal equilibrium | | Condition after UV irradiation | |
|---|---|---|---|---|---|---|
| T (K) | Time (h) | Pressure (mbar) | Time (h) | Pressure (mbar) | Time (h) | Pressure (mbar) |
| 573 | 0 | 15 | 5 | 21 | 23 | 22 |
| 873 | 0 | 16 | 5 | 24 | 23 | 25 |
| 1173 | 0 | 15 | 5 | 29 | 23 | 30 |
| 1273 | 0 | 15 | 6 | 30 | 24 | 31 |
| 1373 | 0 | 16 | 6 | 33 | 24 | 34 |
| 1473 | 0 | 15 | 6 | 33 | 24 | 33 |
| 1473 | 0 | 81 | 6 | 190 | 204 | 150 |

### 2.2. Infrared spectroscopy analysis of the gas phase components

Evolution of the gas mixture in the reaction cell was monitored with a Thermo Scientific Nicolet iG50 Fourier Transform Infrared (FTIR) spectrometer. The path of the FTIR beam in the CAAPSE experiment is presented in Figure 1. A collimated FTIR beam (a few mm in diameter) passes through our high temperature cell and is collected with a $LN_2$-cooled MCT-A detector. In the results presented in Figure 3 and Figure 5, IR spectra are recorded in the 1500-4500 cm$^{-1}$ range with a resolution of 0.25 cm$^{-1}$ after a co-addition of 700 scans. The optical path length inside the cell is 48 ± 1 cm.

## 2.3. Mass spectrometry analysis of the gas phase components

*In situ* measurements of the gas mixture composition in the reaction cell were monitored using a Stanford Research System RGA200 quadrupole mass spectrometer (QMS) equipped with electron-multiplier to increase sensitivity. Gases were transferred to the QMS by opening a high-vacuum leak valve that separates the reaction gas-cell from the pumping system to which the QMS was attached at pressures of ~$1\times10^{-6}$ mbar during the measurements. QMS ionization was achieved through electron impact at 70 eV. The RGA200 covers 1 to 200 *m/z* mass range with a resolution of 100 at *m/z* 100 (*m/Δm*).

## 2.4. Solid phase collection and infrared analysis of thin films

Two sapphire substrates (25 mm diameter and 1 mm thick) were placed inside the CAAPSE cell to collect any aerosols produced during the experiments. Sapphire has a good stability at the studied temperatures. They were positioned at 18 cm and 30 cm from the hydrogen lamp end of the $MgF_2$ window, respectively. The films were deposited at 1473 K during 204 hours of UV irradiation. At the end of the UV-irradiation, the temperature was ramped down to 295 K and the volatiles were pumped off. Subsequently, the cell was opened to ambient air and the samples were transferred for analysis. Single-beam infrared spectra of the samples were measured with a Thermo Scientific Nicolet 6700 Fourier Transform Infrared Spectrometer (FTIR). The infrared signal was collected by a Deuterium TriGlycine Sulfate (DTGS) detector in the 1600 $cm^{-1}$ (sapphire window absorption limit) to 4000 $cm^{-1}$ range with a resolution of 1 $cm^{-1}$ after a co-addition of 300 scans.

## 2.5. Temperature gradient of the cell and gas phase species quantification

We have measured the temperature at different positions of the cell using the three type-B thermocouples distributed across the hottest part of the tube as well as four type-K thermocouples spaced from the center to the ends of the cell, in order to determine the temperature gradient. The temperatures measured with the quartz cell and the alumina cell as a function of tube position for different set-points are presented in Figure 2. Two heating elements distributed across the center of the furnace at a spacing of 10 cm provided heating for the central 10 cm of the furnace. Another 10 cm on each side is thermally insulated and contained in the furnace. As a result, the furnace covered 30 cm of the cell in the middle and the rest of the cell on both the sides (9 cm each) is in thermal contact with the ambient air and chilled water-cooled vacuum flanges near the $MgF_2$ windows.

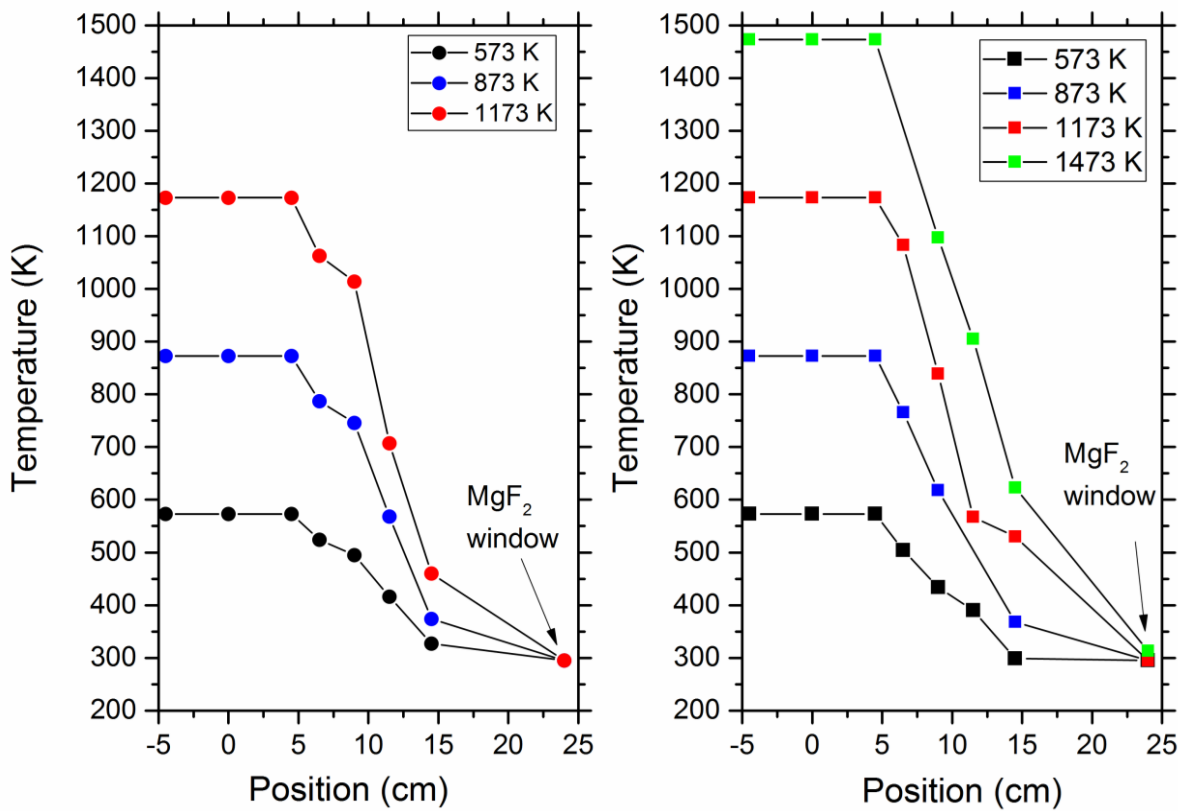

**Figure 2:** Temperature of the cell measured as a function of the cell position for different set oven temperatures with the quartz cell (left) and the alumina cell (right). The origin of the position is set as the center of the cell.

In our experimental setup, the heat is not uniformly distributed along the reaction cell, because of the requirement to keep $MgF_2$ windows at or below 300 °C. As a result, a temperature gradient exists in the gas cell from the center to the end and the gas in the center of the cell is at a higher temperature while the gas at the extremities of the cell (near the $MgF_2$ windows and the UV lamp) is close to 300-400 K. Although the low-temperature gases nearest the UV lamp do absorb some photons, the vast majority of the photons make it to the center of the cell and beyond. Considering CO only, which is the more abundant species in our system after $H_2$, we made an estimate of the percentage of transmission of the photons for these VUV wavelengths (Ly-α at 121.6 nm and 155-165 nm). Using a partial pressure of CO of 0.045 mbar a cross-section (at ambient temperature) ~$4.5\times10^{-20}$ cm$^2$ at 121.6 nm, and ~$2.6\times10^{-18}$ cm$^2$ for the 155-165 nm range, we estimate that ~99.9% of Ly-α photons and ~93.4% of the photons in 155-165 nm range reach the center of the cell where the molecules are at the highest temperature. Taking into account the amounts of $CH_4$ and $CO_2$ (and their cross sections at room temperature) produced by thermal processes before irradiation slightly changes these values to 98.5% for 121.6 nm and 93.4% for 155-165 nm. While these values are only a rough estimate, it shows that the majority of photons will reach the high temperature part of the cell and that the CAAPSE experiment can be used to simulate photochemistry at high temperatures.

To quantify the concentration of a molecule detected in our reaction cell by IR spectroscopy using the Beer-Lambert law, we need to use the absorption cross section of the molecule at the gas temperature. Previous high-temperature studies using similar instrument configurations as ours have measured the emission cross section of $CH_4$ in the infrared (Hargreaves et al. 2012) and the absorption cross section of $CO_2$ in the VUV (Venot, et al. 2018). In both studies, the authors concluded that their spectra were dominated by the emission or absorption of the gas at the maximum temperature in the central portion of the cell. To simplify the analysis of our data, we use the same hypothesis and assume that the majority of the gas is at the maximum temperature $T_{max}$, although it increasing the uncertainties on the values calculated. Then the concentration of a given molecule can be calculated using the Beer-Lambert law. The concentration of the absorbing molecule [$C$] (molecules·cm$^{-3}$) in the cell is defined by *Eq. (1)*:

$$[C] = \frac{\int_{\lambda_1}^{\lambda_2} A \, d\lambda}{l \times \int_{\lambda_1}^{\lambda_2} \sigma \, d\lambda} \qquad (1),$$

where σ is the absorption cross-section (cm$^2$·molecule$^{-1}$) of the molecule at a given wavelength λ and at the maximum temperature in the reaction cell ($T_{max}$), $l$ is the path length (cm) of the beam through the gas cell, and *A* is the absorbance at a given wavelength λ. We have calculated the absorption cross-sections using the HITEMP and ExoMol databases (Rothman et al. 2010; Tennyson et al. 2016). To calculate the concentration [$C$], absorbance A and absorption cross-section σ were integrated between $\lambda_1$ and $\lambda_2$.

## 3. Results and discussions
### 3.1. Thermochemistry: gas-phase composition at the thermal equilibrium

We first investigated the change in the gas phase composition that is induced solely by the heating of the gas mixture (without UV irradiation). Figure 3 presents the IR spectra of the gas mixtures at ambient temperature (reference at 295 K) and after ~2 hours subsequent to ramping to achieve different set maximum temperatures: 573 K, 873 K, and 1173 K, using a quartz cell.

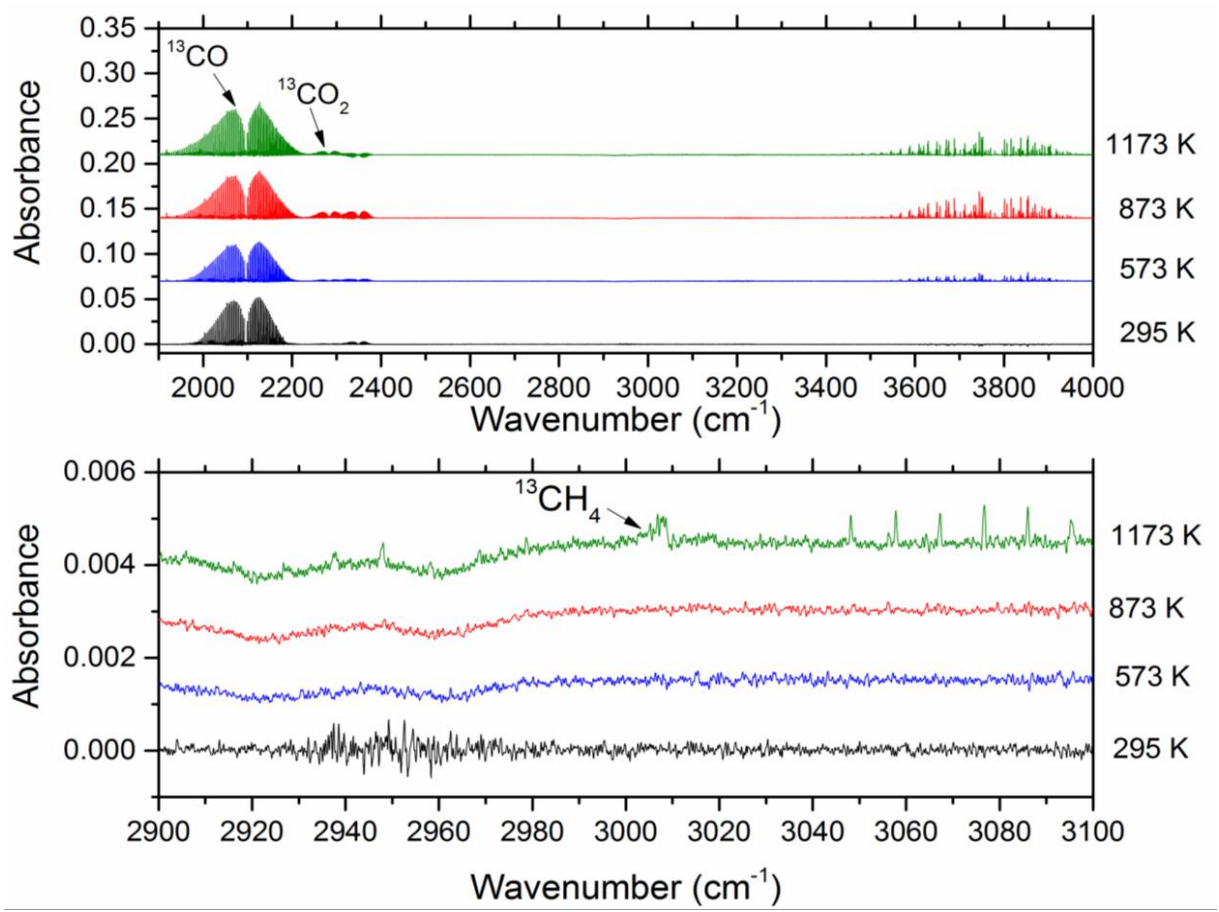

**Figure 3: IR spectra of the initial gaseous mixture of $H_2$:$^{13}CO$ (99.7%:0.3%) at 295 K and at the thermal equilibrium for different set oven temperatures: 573 K, 873 K, and 1173 K. Spectra are offset for clarity. These experiments were conducted using the quartz cell, which is chemically inert and thermally stable only up to ~1300 K.**

The room-temperature spectrum contains the absorption band of $^{13}CO$ centered at 2095 cm$^{-1}$ as well as the absorption band of the residual $^{12}CO_2$ and gas-phase water of ambient air (on the optical pathway outside of the cell) centered at 2348 cm$^{-1}$ and 3750 cm$^{-1}$ respectively. The spectra show that higher $^{13}CO$ rotational states up to J = 43 are thermally populated at 1173 K compared to J = 27 at 295 K. The spectra of the heated gases also reveal the formation of bands due to two new (thermally-generated) molecular species. One band system, centered at 2284 cm$^{-1}$ and visible at the three studied temperatures, is due to $^{13}CO_2$. The other band system centered at 3009 cm$^{-1}$ is visible only in the spectra recorded at 1173 K and is attributed to the Q branch of the υ$_3$ band of $^{13}CH_4$. We also observed several transitions of the P and R branches of $CH_4$ in the 2900-2980 cm$^{-1}$ and 3050-3100 cm$^{-1}$ ranges. These two new species are isotopically labelled with $^{13}C$, highlighting that they are formed during thermal chemistry from the initial $^{13}CO$.

Subsequently, we performed a kinetic study of the formation of $^{13}CO_2$ and $^{13}CH_4$ during the heating to 1173 K in the quartz cell, which is the only temperature where a significant amount of methane formation was observed. Figure 4 presents the evolution of $^{13}CO$, $^{13}CO_2$, and $^{13}CH_4$ concentrations by tracking their absorption band areas as a function of time. For purpose of comparison with photochemical experiments (Section 3.2), we left the gas at the set temperature for 15 hours following ramping. First, we observed that the integrated band area of $^{13}CO$ increased during the heating of the gases before reaching a steady-state. This increase can be explained by a change of the absorption cross-section of $^{13}CO$ with the temperature. Second, we observed that the formation of $^{13}CO_2$ begins within ~50 min of the starting of the heating ramp (at ~600 K), while the formation of $^{13}CH_4$ was seen only after ~ 125 min (corresponding to a cell temperature of 900 K). Within 15 hours of keeping the reaction cell at 1173 K, we observed that concentrations of $^{13}CH_4$ reach a steady-state, demonstrating that an equilibrium exists for this species under these experimental conditions. However, we observed a slow increase of the $^{13}CO_2$ concentration with time, indicating that this species does not reach a steady-state even after 15 hours. After exposure to UV light, the growth of $^{13}CO_2$ appears to accelerate (and $^{13}CO$ decreases). Trends of the evolution of the amounts of these three species over the 15 hours can be fitted using a linear fit as shown in Figure 4.

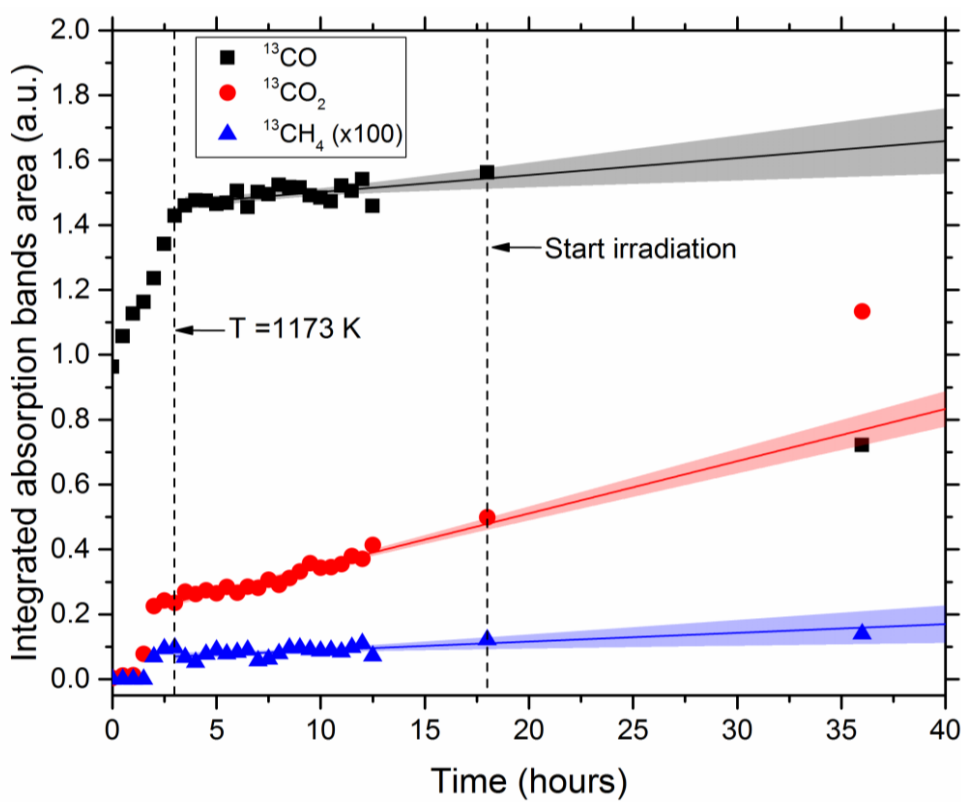

**Figure 4: Evolution of the integrated IR absorption bands of $^{13}CO$, $^{13}CO_2$, and $^{13}CH_4$ as a function of time. The origin of the time is set at the moment when the furnace heating**

ramp started. The first vertical dashed line marks the moment where the furnace temperature reaches 1173 K and the second vertical dash line marks the beginning of the irradiation. Linear fits of evolution of the integrated absorption bands of the three species over 15 hours at 1173 K are also displayed with 95 % confidence intervals.

We quantified the mixing ratios of $^{13}CO$, $^{13}CO_2$ and $^{13}CH_4$ at the thermal equilibrium using the method described in Section 2.5 and these ratios at different temperatures are summarized in Table 2. The initial mixing ratio of $^{13}CO$ varied slightly from one experiment to the next, due to a slow escape of $H_2$ from the storage glass flask containing the premixed gases at ~500 mbar, making the second experiment to have a slightly higher $CO/H_2$ ratio. Though $^{13}CO$ remained the main carbonaceous species at different equilibrium temperatures, meaning that only a minor part of $^{13}CO$ underwent thermochemistry in the cell, the concentration of $^{13}CO_2$ increased with temperature, reached a maximum, and subsequently decreased at elevated temperatures (Table 2). Finally, methane production was observed only at 1173 K with an equilibrium mixing ratio of ~80 ppm$_v$. If $CH_4$ was produced at 573 K and 873 K, its abundance was below the limit of detection of our FTIR. At higher temperatures, $CH_4$ may be undergoing further thermal chemistry as we have not detected $CH_4$ at the other temperatures listed in Table 2.

**Table 2: Mixing ratios of $^{13}CO$, $^{13}CO_2$, and $^{13}CH_4$ calculated at the thermal equilibrium composition for different set oven temperatures using the quartz and the alumina cells. The uncertainties are given at 2σ (standard deviation) and were calculated from the standard fluctuations of the infrared spectroscopy measurements.**

| T (K) | Reaction cell | $^{13}CO$ | $^{13}CO_2$ | $^{13}CH_4$ |
|---|---|---|---|---|
| 295 | Quartz | $(3.8 \pm 0.2)\times10^{-3}$ | - | - |
|  | Alumina | $(3.4 \pm 0.2)\times10^{-3}$ | - | - |
| 573 | Quartz | $(4.0 \pm 0.2)\times10^{-3}$ | $(1.5 \pm 0.1)\times10^{-5}$ | - |
|  | Alumina | $(4.1 \pm 0.2)\times10^{-3}$ | $(1.0 \pm 0.1)\times10^{-5}$ | - |
| 873 | Quartz | $(4.8 \pm 0.3)\times10^{-3}$ | $(6.4 \pm 0.4)\times10^{-5}$ | - |
|  | Alumina | $(4.1 \pm 0.2)\times10^{-3}$ | $(1.6 \pm 0.1)\times10^{-5}$ | $(4.3 \pm 0.9)\times10^{-5}$ |
| 1173 | Quartz | $(5.5 \pm 0.3)\times10^{-3}$ | $(3.8 \pm 0.2)\times10^{-5}$ | $(8.0 \pm 0.8)\times10^{-5}$ |
|  | Alumina | $(4.7 \pm 0.3)\times10^{-3}$ | $(1.4 \pm 0.1)\times10^{-4}$ | - |
| 1273 | Alumina | $(5.0 \pm 0.3)\times10^{-3}$ | $(1.8 \pm 0.2)\times10^{-4}$ | - |
| 1373 | Alumina | $(5.3 \pm 0.3)\times10^{-3}$ | $(8.0 \pm 0.4)\times10^{-5}$ | - |
| 1473 | Alumina | $(5.6 \pm 0.3)\times10^{-3}$ | $(3.4 \pm 0.2)\times10^{-5}$ | - |

Thermal chemistry in exoplanet atmospheres is already studied using chemical models, with a particular focus on reactions converting oxidized carbon to reduced carbon such as the CO-CH$_4$ quenching reactions, see e.g. (Moses, et al. 2011; Visscher & Moses 2011; Moses 2014). However, gas-phase reactions considered in these models include reaction of CO with atomic hydrogen, which may not be a major component in our work conditions. Although we did not identify chemical pathways for the thermal generation of CO$_2$ and CH$_4$ in our experiments, it has been shown experimentally that mineral phases could efficiently catalyze H$_2$:CO thermochemistry at temperatures up to 900 K (Hill & Nuth 2003), and it is likely that reactions catalyzed by the wall of the cell may play an important role in the thermochemistry observed in our experiments. Further experiments would be necessary to verify this theory but it is beyond the scope of this work.

To extend our experiments to the higher temperatures that exoplanet atmospheres may experience, we used another cell made of alumina instead of quartz. Alumina has a better resistance to higher temperatures but alumina is more porous than quartz, enhancing the surface area of gas-solid contact significantly and possibly also catalytic effects, increasing overall the thermochemistry efficiency. To evaluate this effect, we reproduced the experiments done with the quartz cell and extended our study to higher temperatures.

Figure 5 presents the IR spectra of the gas mixture at ambient temperature (reference at 295 K) and after the heating of the gases at different set temperatures of the alumina cell: 573 K, 873 K, 1173 K, 1273 K, 1373 K, and 1473 K. Similar to the experiments conducted with the quartz cell (Figure 3), we observed heating of the gases (indicated by higher rotational states of $^{13}$CO that were thermally populated at high temperatures), along with the formation of $^{13}$CO$_2$. Similar to the observations with quartz cell, the concentration of $^{13}$CO$_2$ increased with temperature, reached a maximum, and then decreased at higher temperatures (Table 2). The $^{13}$CO$_2$ production was maximized at 873 K with the quartz cell while it was maximized at 1273 K with the alumina cell.

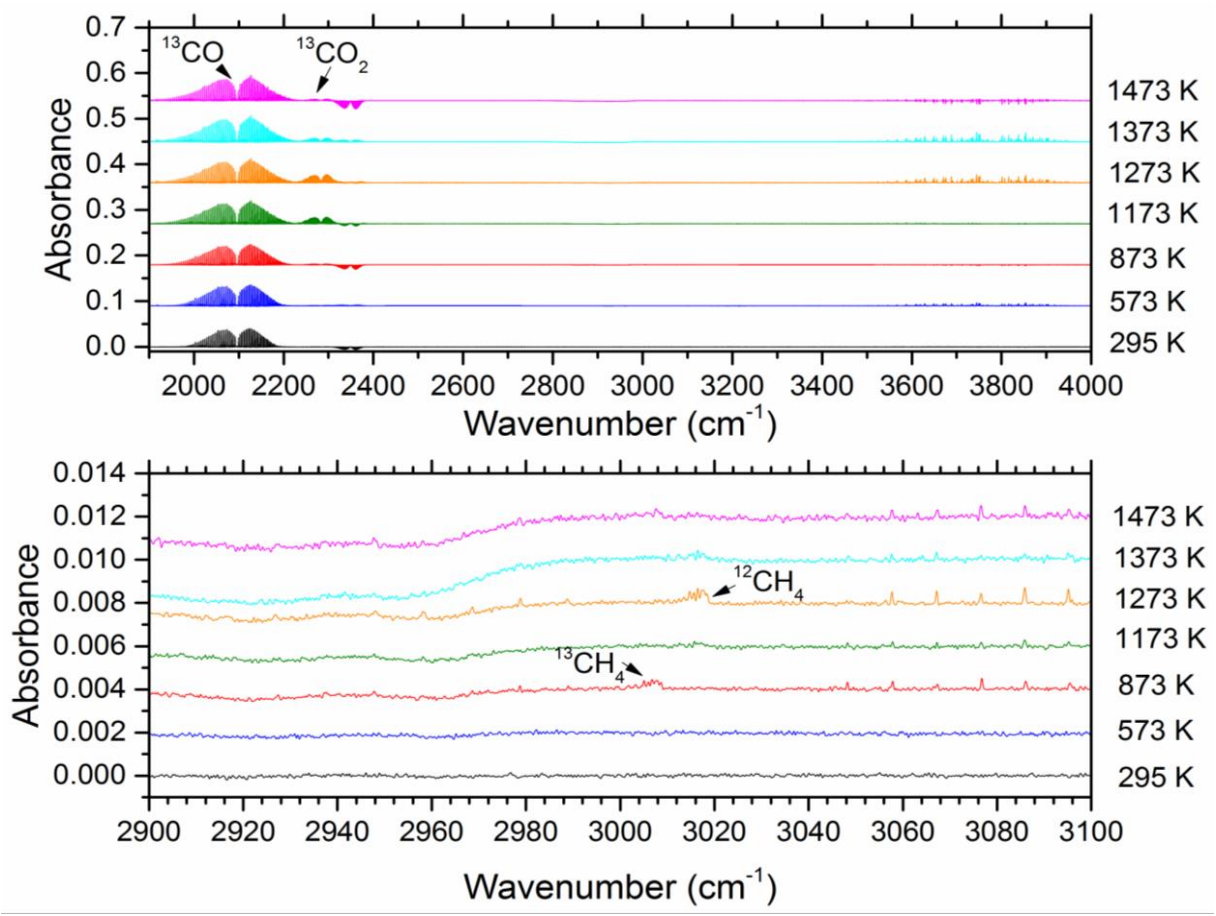

**Figure 5:** IR spectra of the initial gaseous mixture of $H_2$:$^{13}CO$ (99.7%:0.3%) at 295 K and at the equilibrium for different set oven temperatures: 573 K, 873 K, 1173 K, 1273 K, 1373 K, and 1473 K; spectra are offset for clarity. The experiments were conducted using the alumina cell for this dataset.

Formation of $^{13}CH_4$ was observed in small quantities at a much lower temperature (873 K) when the alumina cell was used (Figure 5), in contrast to its detection at 1173 K when the quartz cell was used (Figure 3). It is likely that the alumina tube acts as a much more efficient catalyst with larger surface area generating and perhaps consuming methane further. It should be noted that previous studies have shown that a decrease of the transition intensity of the $\upsilon_3$ absorption band of $CH_4$ and an increase of the band continuum (Alrefae et al. 2014; Hargreaves et al. 2015) when the gas temperature increases. Therefore, the signal-to-noise ratio of the absorption band of $CH_4$ should decrease drastically at higher temperatures and may lead to a decrease of the detection sensitivity of $CH_4$ with our spectrometer. Overall, we find that in general $CH_4$ is not the major component in the thermochemical equilibrium composition. We also have observed $^{12}CH_4$ (3017 cm$^{-1}$) in the reaction cell above 1173 K, which should be due to two potential sources of $^{12}C$ isotope products discussed in Section 3.2.3.

The presence of carbon dioxide and methane is predicted by models for hot, carbon-rich, hydrogen-dominated exoplanet atmospheres such as those simulated in this work. However, the final observed mixing ratios observed here do not reproduce the ones calculated in chemical models. In particular, if a methane mixing ratio of few tens of ppm$_v$ remains consistent with model calculations, carbon dioxide is predicted to be present at an abundance of few ppb, orders of magnitude lower than was measured in our experiments (Venot, et al. 2015; Heng & Lyons 2016). This discrepancy between thermochemical models and our laboratory experiments in regard to the formation of $CO_2$ needs to be evaluated in future work.

### 3.2. Photochemistry of high temperature mixtures at thermal equilibrium

In UV-rich stellar environments, photochemistry would affect exoplanet atmospheres that are at hot gas temperatures and their compositions would depart from thermal equilibria. VUV ($\lambda < 200$ nm) photons are the main energy source driving photochemistry in these environments. Among them, hydrogen emission at 121.6 nm (Ly-$\alpha$) has an important contribution to photochemistry, particularly for exoplanets orbiting M-type stars (France, et al. 2013; Miguel, et al. 2015). Although all VUV photons contribute to photochemistry, irradiation with Ly-$\alpha$ photons is a good way to reproduce photochemistry driven by stellar photons. However, the energy of Ly-$\alpha$ photons (10.2 eV) is not sufficient to dissociate or ionize either $H_2$ or CO. Hence, photochemistry can only be initiated with electronic excitation of CO at the wavelengths emitted by the hydrogen lamp. However, both $CO_2$ and $CH_4$ can be directly photodissociated under these conditions and their absorption cross-section is temperature dependent (Chen & Wu 2004; Venot, et al. 2018).

From our thermochemistry study presented in Section 3.1 it is clear that, despite some quantitative differences, the heating of the gases in the quartz cell and in the alumina cell resulted in the production of similar species at thermal equilibria. For this reason and for the reason that we can access higher temperatures, we chose to conduct UV-photochemistry experiments using the alumina cell.

After attaining thermal equilibria (Figure 5), the same gas mixtures were irradiated for a duration of 18 hours each in the alumina cell equipped with $MgF_2$ windows. Figure 6 presents the IR spectra of the gas mixtures after 18 hours of irradiation for different set oven temperatures: 573 K, 873 K, 1173 K, 1273 K, 1373 K, and 1473 K.

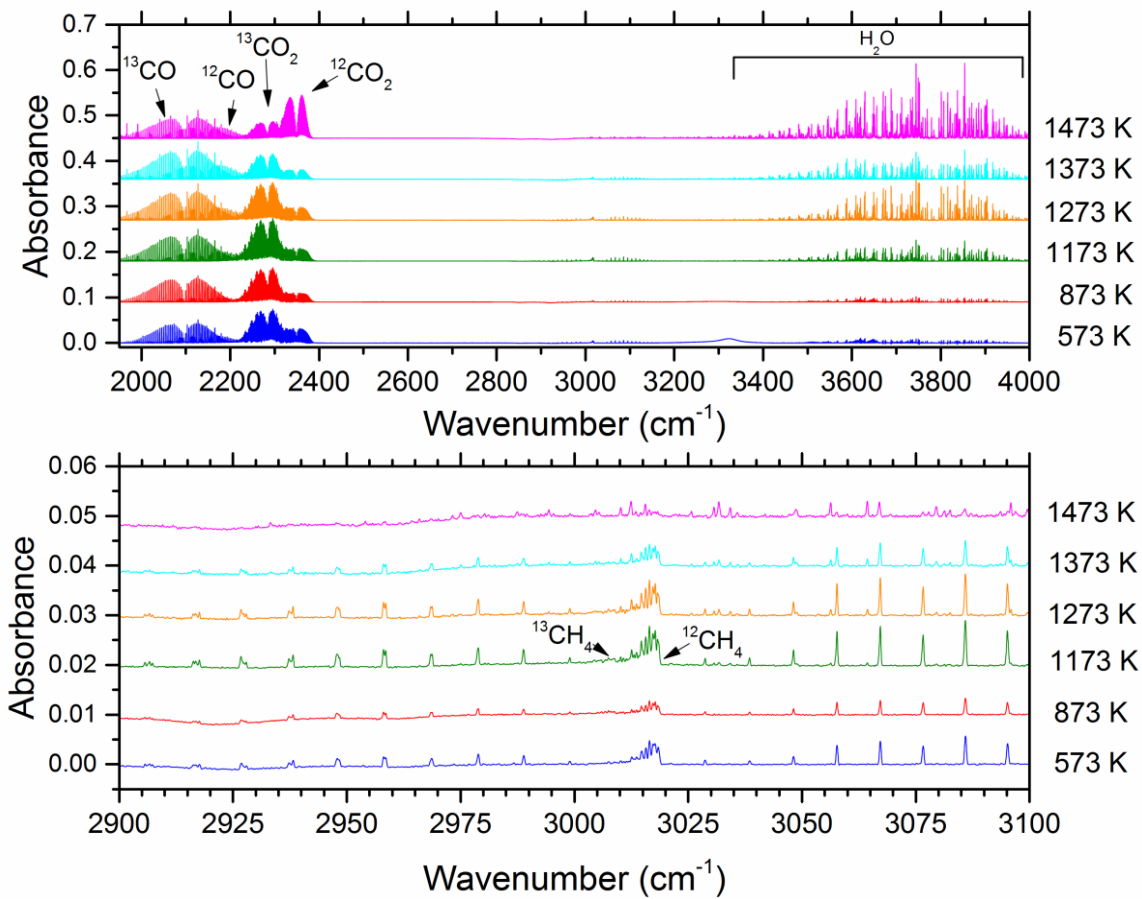

**Figure 6 :** IR spectra of the gaseous mixture (beginning ratio of $H_2$:$^{13}CO$ = 99.7%:0.3%) after 18 hours of irradiations at 121.6 nm (Ly-α) for different set oven temperatures: 573 K, 873 K, 1173 K, 1273 K, 1373 K, and 1473 K. Experiments were realized with the alumina cell.

An artefact is visible at ~3320 cm$^{-1}$ on the spectrum recorded for a set temperature of 573 K. The spectra contain the absorption bands of carbon monoxide ($^{13}CO$ and $^{12}CO$), carbon dioxide ($^{13}CO_2$ and $^{12}CO_2$), methane ($^{13}CH_4$ and $^{12}CH_4$), and $H_2O$. Additionally, three weak absorption bands have been observed, centered at 3527 cm$^{-1}$, 3631 cm$^{-1}$, and 3715 cm$^{-1}$. These bands are convoluted with the signature of $H_2O$ and are observed only for temperatures lower than 1200 K. However, we have not been able to identify the corresponding species. The amount of $^{13}CO$ decreased after the irradiation, and the consumption efficiency increased with the temperature, which highlights an efficient photochemistry of $^{13}CO$. However, since $^{12}CO$ and $^{13}CO$ absorption bands overlap, we could not perform a quantification of $^{13}CO$.

### 3.2.1. Carbon dioxide and water formation

We quantified the mixing ratio of $^{13}CO_2$ after 18 hours of Ly-α irradiation at each studied temperature. The results are presented in Table 3. For all the temperatures, the mixing ratio of $^{13}CO_2$ increased by one or two orders of magnitude after the irradiation compared to the values measured at the thermal equilibria (Table 2). Although $^{13}CO_2$ concentration does not reach a steady-state over several hours in thermal-only conditions, its production rate is too low to explain the large amount of $^{13}CO_2$ observed after the irradiation as illustrated in Figure 4 for an experiment made at 1173 K with a quartz cell. However, this effect must be taken into account to interpret quantitative evolution of the carbon dioxide concentration.

**Table 3: Mixing ratios of $^{13}CO_2$ (with respect to $H_2$) calculated after 18 hours of Ly-α irradiation for different set oven temperatures: 573 K, 873 K, 1173 K, 1273 K, 1373 K, and 1473 K. The uncertainties are given at ±2σ (standard deviation) and are calculated from the standard fluctuations of the infrared spectroscopy measurements.**

| T (K) | $^{13}CO_2$ |
|---|---|
| 573 | $(1.0 \pm 0.1) \times 10^{-3}$ |
| 873 | $(8.5 \pm 0.1) \times 10^{-4}$ |
| 1173 | $(1.2 \pm 0.1) \times 10^{-3}$ |
| 1273 | $(1.0 \pm 0.1) \times 10^{-3}$ |
| 1373 | $(6.4 \pm 0.1) \times 10^{-4}$ |
| 1473 | $(4.3 \pm 0.1) \times 10^{-4}$ |

The production of $CO_2$ through the photoexcitation of pure CO has been experimentally studied with irradiation at 123.58 nm and 206.2 nm (Dunn, et al. 1973; Vikis 1982). This leads to the excitation of CO at the $A^1\Pi$ and $A^3\Pi$ states, respectively.

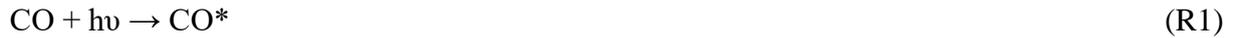

CO + hυ → CO* (R1)

Then, the excited molecules react with ground state CO to form primarily $CO_2$ but also a small amount of $C_3O_2$, which is not observed in our experiments (Liuti, et al. 1966b, 1966a).

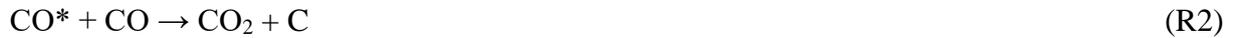

CO* + CO → $CO_2$ + C (R2)

The excitation of $^{13}CO$ by the Ly-α radiation (121.6 nm) is responsible for initiating the generation of $^{13}CO_2$ in our experiments. The concentration of $^{13}CO_2$ first increased with the temperature to a maximum at 1173 K before decreasing at higher temperatures. Although $CO_2$ can be efficiently formed by photochemistry, it can be also thermally decomposed and photodissociated by UV photons. The efficiencies of these two loss processes increase with temperature (Venot, et al. 2018) and could explain the lower $CO_2$ concentration at higher temperatures.

In addition, a net production of water was observed despite a competitive destruction by UV photons. The production of water generally increased with the temperature according to the spectra in Figure 6. However, it should be noted that these spectra may contain a minor contribution from the residual atmospheric water on the optical pathway outside of the reaction cell. An efficient water formation pathway has previously been reported in experiments conducted at room temperature in $N_2/CO_2/H_2$ gas mixtures using a plasma discharge as energy source (Fleury et al. 2015). The formation of water can be initiated by the formation of $O(^1D)$ radicals by photodissociation of $CO_2$ in our experiments. The photodissociation of $CO_2$ can follow two different pathways, depending on the photon energy:

$$CO_2 + h\upsilon \rightarrow CO + O(^3P) \tag{R3}$$

$$CO_2 + h\upsilon \rightarrow CO + O(^1D) \tag{R4}$$

At wavelengths shorter than 167 nm, reaction R4 is dominant (Huebner et al. 1992) and so the photodissociation of $CO_2$ principally leads to the formation of $O(^1D)$ radicals, which can then react with $H_2$ to form OH radicals (Vranckx et al. 2010):

$$H_2 + O(^1D) \rightarrow OH + H \tag{R5}$$

Finally, the formation of water is explained through the following reactions (Orkin et al. 2006; Bahng & Macdonald 2007):

$$H_2 + OH \rightarrow H_2O + H \tag{R6}$$

$$2OH \rightarrow H_2O + O(^3P) \tag{R7}$$

The increase of the water production with temperature is correlated with the decrease of the carbon dioxide mixing ratio. Experimental measurements have shown that the $CO_2$ absorption cross-section can increase by orders of magnitude as the temperature is increased (Venot, et al. 2018). Thus, an increase in temperature could lead to more efficient photodissociation of $CO_2$, leading to higher production of $O(^1D)$ radicals and water despite the competitive water photodissociation process.

Here, we have shown that photochemistry can strongly influence the composition of $H_2$:CO warm exoplanet atmospheres. In particular, our results demonstrate that carbon dioxide and water can be efficiently produced in these atmospheres despite competitive loss processes. We find that the gas temperature strongly influences the efficiency of the different chemical pathways. This is notably due to the increase of the absorption cross-sections of $CO_2$ when the temperature increases (Venot, et al. 2018). A similar effect is expected for other molecules such as CO and $H_2O$ and more laboratory measurements are needed to address this further. Moreover, chemical reaction kinetics is known to be strongly affected by the temperature.

### 3.2.2. Volatile organics formation

Electronically-excited CO (equation R1) can also react with the excess of $H_2$ molecules available in the chamber or H atoms generated through thermal/photochemical dissociation of $H_2$ or water, initiating the hydrogenation of CO (forming HCO) (Roussel & Back 1990) that could eventually result in the formation of formaldehyde ($H_2CO$) or methane ($CH_4$). Alternatively, carbon formed by equation R2 can react with $H_2$ or H or CO, initiating the formation of the precursors CH and $C_2O$, leading to the formation of complex organics as well as methane. Literature studies determined that in the ground-state CO + H reaction has a barrier of about 4 kcal/mol ((Pham et al. 2016) and references therein) and concluded that HCO formation must occur in CO* excited state, leading all the way to the formation of $H_2CO$ and $CH_3OH$ (not observed in our experiments).

### 3.2.3. Isotopic Labeling and Natural Isotope Products

Isotopically-enriched $^{13}CO$ has been used in all the experiments reported here, in order to differentiate any potential leaks into the reaction cell which was at lower pressure (~15 – 30 mbar). We detected normal isotope $^{12}C$ products during thermochemistry phase in alumina reaction cell ($^{12}CH_4$), as well as during the photochemistry phase ($^{12}CH_4$, $^{12}CO_2$ and $^{12}CO$). We could envision two potential sources of normal isotope products in our experiments: (1) an extremely small leak of ambient air into the reaction cell, (2) pyrolysis and photolysis of organic residue (most likely from the solvents such as isopropanol used to clean) on the inner walls of the cell. Quartz has a much smoother surface than the alumina, hence there is a larger inner-wall surface area for the latter. $^{12}CH_4$ is observed only during the thermal experiments with alumina cell at 1173 K and higher temperatures (Figure 5), but not when the quartz cell was used at <1173 K (Figure 3). This could most likely be due to pyrolysis of organic residue (scenario 2), because we do not detect any increase of $^{12}CO_2$ or $^{12}CO$ in the reaction cell during thermochemistry studies. Venot et al. (2018) performed studies on $CO_2$ spectroscopy at higher temperature (similar to our conditions) and found that CO was formed by thermal decomposition of $CO_2$. If a small leak of ambient air into the cell (scenario 1) were to be present in our experiment, we would expect $^{12}CO_2$ from the ambient air to contribute to an increase of

$^{12}CO_2$ spectral bands as well as thermal dissociation of $^{12}CO_2$ from the leak to increase the $^{12}CO$ spectral bands. So, we can confidently rule out a leak during the thermochemistry phase of our experiments. During the photochemistry phase, we detect significant enhancement of $^{12}CO_2$ signals, followed by $^{12}CO$ and $^{12}CH_4$. This could be due to pyrolysis and photolysis of the organic residue (scenario 2).

Further support for the possibility of photolysis of residual organics from the walls of the cell comes from our mass spectrometry measurements that are complementary to the infrared spectroscopic measurements. We also analyzed the gas-phase volatiles of the cell *in situ* by mass spectrometry during these experiments before and after thermal equilibrium as well as after photolysis. Due to the fact that all the components are undergoing chemical modifications continuously, we took the key component $^{13}CO$ as the molecule that is normalized in each mass spectrum and determined the relative abundances of the rest of the species as shown in Figure 7. However, it is also possible that the absolute concentrations of other molecules might not have changed, while their apparent increase or decrease might be caused by the changing absolute concentration of $^{13}CO$, which we used to normalize the mass spectra. In a system where all the species change in their concentration, it is difficult to obtain absolute quantification and the only possibility is relative quantification. In the future, we may use a rare-gas such as He (which is present in hot-Jupiter atmosphere), Ne, or Ar as a tracer to determine absolute concentrations from mass spectrometry with the assumption that these rare-gas atoms are chemically inert and do not undergo chemical reactions with the gas mixture. Please note that the intensity scales are different for each temperature.

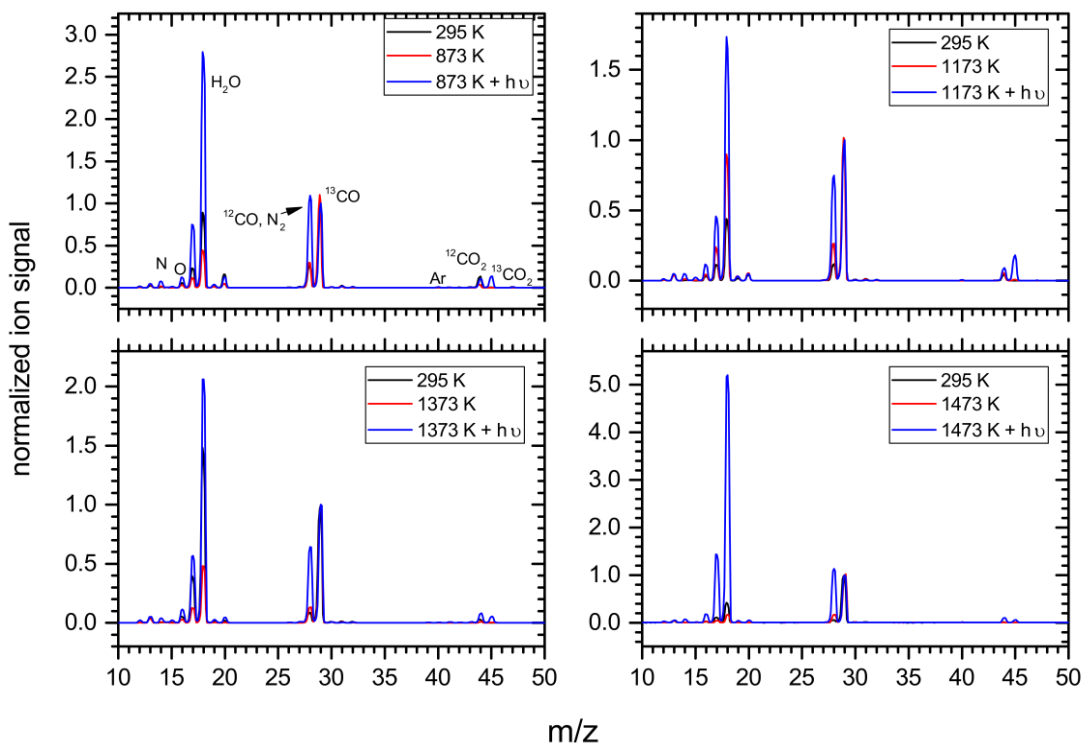

**Figure 7:** Mass spectra of the gas-phase composition of the reaction cell, introduced into the mass spectrometer (using an ultrahigh-vacuum leak valve) and normalized to $^{13}CO^+$ peak at *m/z* 29. Red and Blue curves are before and after UV-photolysis at a given temperature. We see clear enhancement of $H_2O$ and $^{13}CO_2$ as well as the $^{12}CO^+ + N_2^+$ peak at *m/z* 28 upon photolysis.

We find the general tendency of increasing production of $H_2O$ with temperature, and an increase and then decrease of $^{13}CO_2$ with temperature, in agreement with Table 2. We also notice some atmospheric leaks, clearly represented by $N^+$ (*m/z* 14), but also with some minor contribution from $CH_2^+$. However, the peak at 28 could be due to $N_2^+$ as well as $^{12}CO^+$. We monitored the pressure change in the reaction cell and found that the change after 18 hours was only 1 mbar between the beginning of UV-irradiation (after thermal equilibration) and the end of irradiation (Table 1). If increase in pressure of ~ 1 mbar were to be due entirely to $N_2$ (80% air), we would have expected the signal at *m/z* 28 to be at least an order of magnitude larger than the $^{13}CO^+$ at *m/z* 29, which was not the case. The peak at *m/z* 28 ($^{12}CO^+$ and $N_2^+$) is at the maximum about the same height as the peak at *m/z* 29 ($^{13}CO^+$) and no significant $O_2^+$ was detected at *m/z* 32. Based on these qualitative observations and due to the fact that in the FTIR spectrum significant increase in $^{12}CO$ and $^{12}CO_2$ has been observed subsequent to the photochemistry phase, we think that the observed spectra of normal isotope molecules are likely due to pyrolysis followed by photochemically-transformed organic residue within the interior walls of the alumina reaction cell. An atmospheric gas leak, though it cannot be completely ruled out, could only be a minor contributor for the normal isotope molecules.

To better constrain the contribution from outgassing/pyrolysis from the walls of the alumina cell during our experiments and to better understand how they affect our experimental results, we performed an additional control experiment. The alumina cell was heated and pumped down to $3\times10^{-7}$ mbar (Section 2.1), isolated from vacuum, and heated for 24 hours at 1473 K. The composition of the gases inside of the reaction cell was analyzed using IR spectroscopy and mass spectrometry. The mass spectrum at room temperature as well as the mass spectrum and IR spectrum after 24 hours of heating at 1473 K are presented in Figure 8. These spectra show the presence of $H_2O$, $CO_2$ and at a lesser extent $^{12}CO$ (convoluted with water in the IR spectrum) after the heating of the gases. These results are in agreement with the contamination observed after the photochemical experiments. Desorbed water intensities during this control study are 3 times lower than those observed from the photolysis of $H_2$:CO mixture at 1473 K. We also note that the pressure in the sealed alumina cell increased from $3\times10^{-7}$ mbar initially to ~ 4 mbar after 24 hours at 1473 K in this control experiment. For photolysis experiment under the same temperature, but at higher CO/$H_2$ gas starting pressure, we did not notice any change in the pressure between before and after 18 hours photolysis (initial: 33 mbar, final: 33 mbar, detection limit < 0.5 mbar, which is the resolution of our pressure gauge display). Increase of the pressure in the cell during the control experiment is explained by the outgassing at higher temperature of water and $CO_2$ adsorbed on the cell walls. Though similar process is expected to occur during the photolysis experiment at 33 mbar pressure in the cell, both desorption and re-adsorption efficiencies could vary at this relatively higher pressure and compete with each other. Further, photochemistry also produces solid aerosols, which should contribute to net decrease of the pressure if there were no thermal desorption from the cell walls. Although we are unable to definitively quantify water production in our photolysis experiments, these data put together suggest that the water production observed during the photochemistry experiments is significant and that water originating from ambient contamination is minimal.

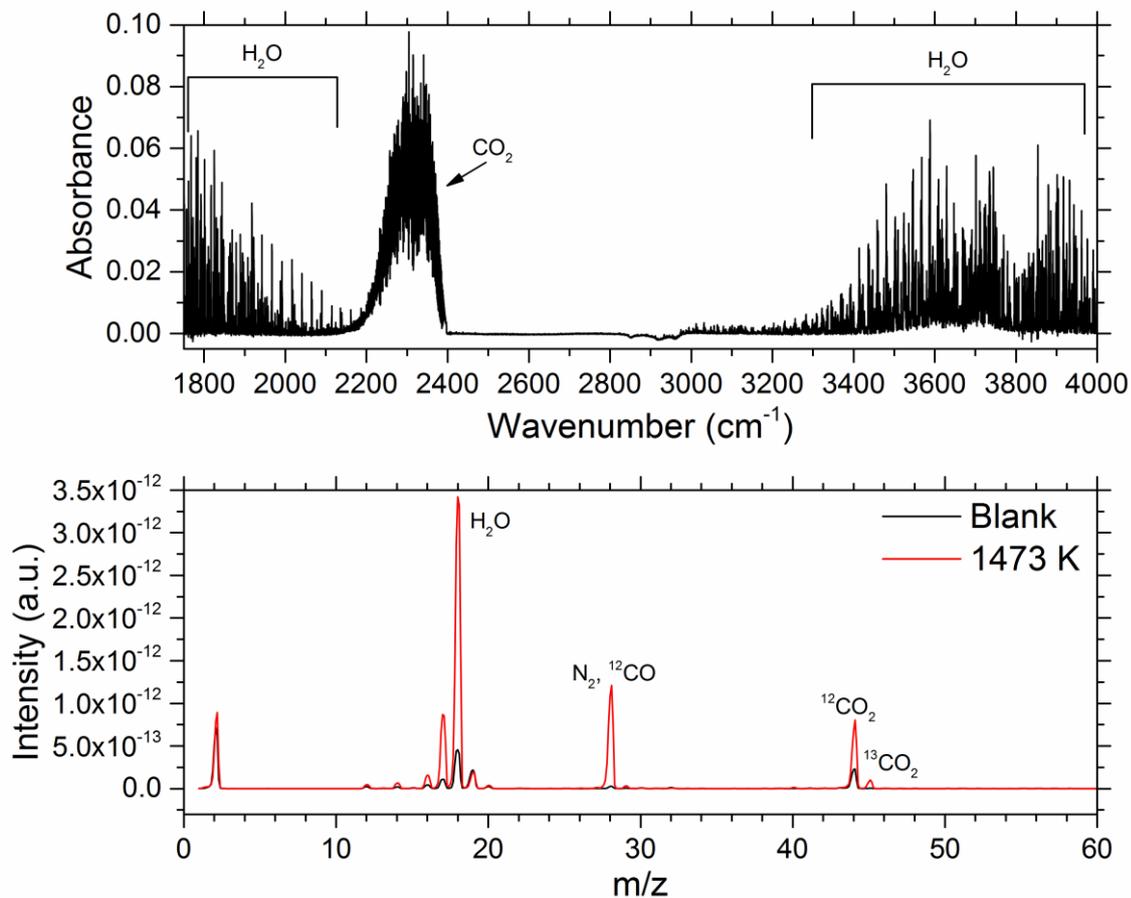

**Figure 8: Spectra from a control experiment whereby the furnace was first evacuated to $3\times10^{-7}$ mbar, then isolated and heated to 1473 K for 24 hrs. Infrared spectrum (top) and mass spectra (bottom) of the gas-phase inside the reaction cell before and after heating at 1473 K for 24 hours.**

Our study, taking the $^{13}CO$ as the initial reactant demonstrates the importance of using such isotopes to unambiguously distinguish various pathways that could potentially interfere with the observations and derived quantitative data points. Though we observed a significant amount of water formation, we could not quantify the mixing ratio due to ambient $H_2O$ in the FTIR optical pathway outside the cell and possible contamination. We want to resolve this problem in the future by using $^{13}C^{18}O$ and $D_2$ to quantify the mixing ratio of $H_2O$ generated through photochemistry of $H_2$ and CO at hot-Jupiter atmospheric conditions.

## 3.3. High temperature photochemical formation of solid refractory organics

A detectable amount of solid phase product was not observed in the experiments presented above. However, we noted that depletion of CO was higher when irradiated at higher temperatures, while the production efficiency of the gas-phase products ($CO_2$ and $CH_4$) decreased, meaning another chemical reaction pathway becomes efficient at higher temperatures under exposure to UV-photons. We propose that higher temperatures enhance the conversion of CO, $CO_2$, and $CH_4$ to solid-phase aerosols. To investigate this hypothesis, we repeated the UV irradiation of $H_2$/CO gas mixture at the thermal equilibrium at 1473 K. The gas mixture was irradiated for 204 hours in order to ensure that a sufficient amount of solid was produced. For this experiment, we placed two sapphire windows inside the cell inside the highest temperature zone at 18 cm and at 30 cm from the UV source. We also increased the pressure in the cell to 81 mbar in order to increase the amount of $^{13}$CO (keeping the same $H_2$:CO ratio as the other experiments) that could be converted into solid organic product.

At the thermal equilibrium, we observed an increase of the pressure to 190 mbar (Table 1) due to the thermal expansion of the gases. Then, after the irradiation, the pressure in the cell had dropped to 150 mbar, highlighting a conversion of a part of the gas mixture into a solid product. Figure 9 presents the photographs of a clean sapphire window (blank) and of the two samples collected after the UV irradiation. We observed a white deposit on the two sapphire windows placed inside the cell. The film deposited on Sample 1 is thicker than the one deposited on Sample 2, which was placed further away from the UV source, but at the same temperature. This observation can be rationalized on the basis of higher attenuation of Ly-α photons at farther distances in the cell away from the UV source. As a result, Sample 2 witnessed lesser photochemical aerosol formation than Sample 1.

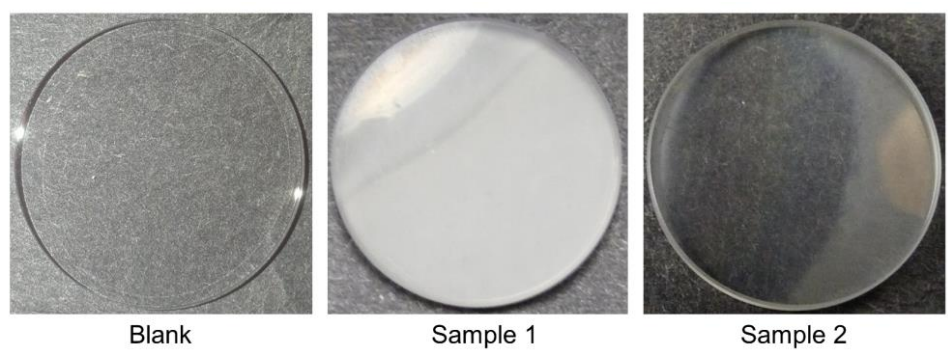

**Figure 9: Photographs of solid-phase organic aerosol products deposited on sapphire windows (1" diameter) after 204 hours of $H_2$:$^{13}$CO (99.7%:0.3%) gas mixture irradiation**

**at Ly-α (121.6 nm) and 1473 K. Blank is a clean sapphire window. Sample 1 and 2 were deposited on two sapphire ($Al_2O_3$) windows placed in the CAAPSE cell at 18 cm and 30 cm from the hydrogen lamp, respectively. Temperatures at Sample 1 and Sample 2 were approximately the same (~1300 K) but more UV photons reached Sample 1 than Sample 2, due to the attenuation of Ly-α (121.6 nm) by CO molecules.**

The thin films formed in our experiment are organics as confirmed by their mid-IR signature presented in Figure 10. We observed that the absorbance of Sample 2 is lower than that of Sample 1, in agreement with their difference of thickness. Different absorption bands characteristic of solid organics are observed in these spectra.

The absorption band centered at 1772 $cm^{-1}$ can be attributed to a C=O carbonyl stretch. Three absorption bands are also observed at 2775 $cm^{-1}$, 2841 $cm^{-1}$ and 2875 $cm^{-1}$. The band at 2775 $cm^{-1}$ can be attributed to the aldehyde –CH stretching mode (H-(C=O)-R). The 2841 $cm^{-1}$ and 2875 $cm^{-1}$ bands are attributed to the –$CH_2$ and –$CH_3$ symmetric and asymmetric stretching modes. The spectrum of Sample 1 shows a broad band centered at 3100 $cm^{-1}$, which could correspond to the aromatic –CH stretching mode. Finally, the most intense absorption band is centered at 2453 $cm^{-1}$. Some inorganic carbonates such as calcium carbonate ($CaCO_3$), sodium carbonate ($Na_2CO_3$) or potassium carbonate ($K_2CO_3$) have combination bands at ~2500 $cm^{-1}$, while much stronger fundamental transitions are located at $υ < 1500$ $cm^{-1}$ (Miller & Wilkins 1952). The sapphire substrate used in our experiments is not transparent at $υ < 1500$ $cm^{-1}$ and new experiments would be necessary to confirm possibility of this band to be due to carbonates. The gas-phase HCO radical has also strong absorption at 2434 $cm^{-1}$ (Jacox 2004). Although its formation from $H_2$ and CO would not be surprising (Pham, et al. 2016), its persistence at 1473 K in the chamber and subsequently on the sapphire window at room temperature is unlikely. One explanation is that HCO could be etched into the sapphire window and trapped through chemisorption. Further work is necessary to understand the nature of this 2453 $cm^{-1}$ band and the exact nature of the species that it is associated with.

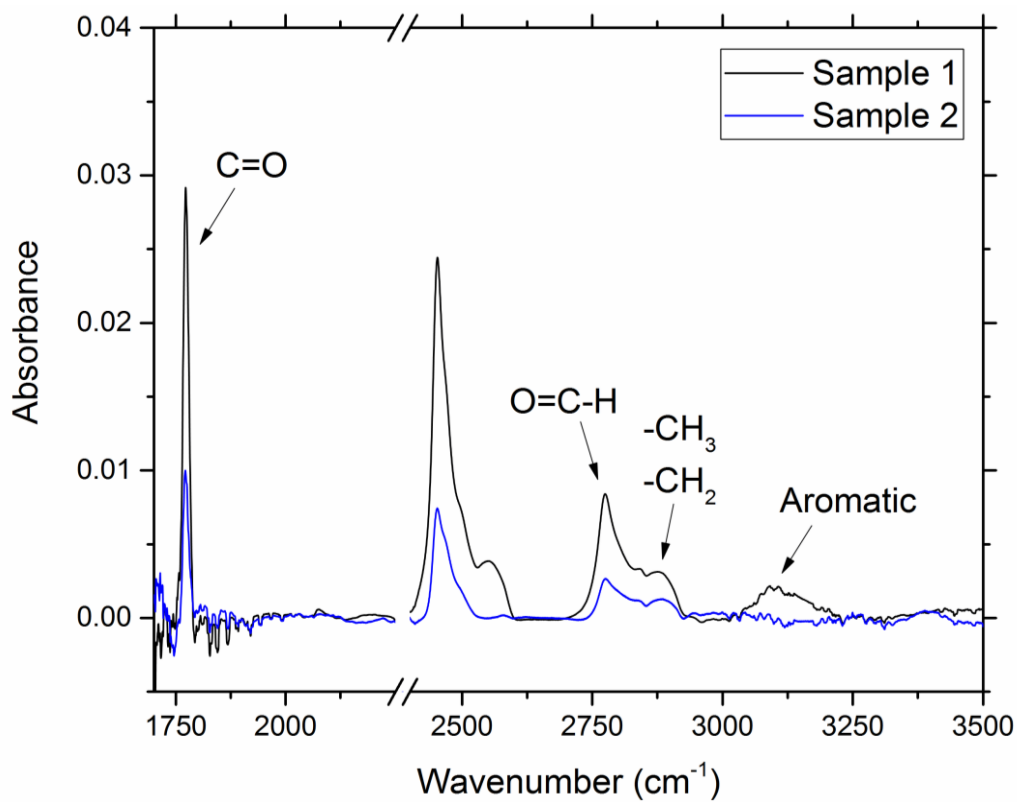

**Figure 10:** Infrared absorption spectra of the films deposited on two sapphire substrates during the same experiments of irradiation of $H_2$:CO (99.7%:0.3%) gas mixture at the thermal equilibrium at 1473 K with Ly-α (121.6 nm) photons. Sample 1 and 2 were deposited on two sapphire windows placed in the CAAPSE cell at 18 cm and 30 cm from the hydrogen lamp respectively.

The formation of a solid-phase material made of C, O, and H in our experiment demonstrates that refractory organic aerosols may be formed in hot exoplanet atmospheres with a C/O ratio of 1, representing C enhancement relative to solar values. This critical result demonstrates the importance of laboratory investigations of the formation and properties of aerosols under simulated exoplanet atmospheric conditions, in order to improve the characterization of these atmospheres.

### 3.4. Discussion

Our experimental results provide new and first insights into the photochemical processes occurring in hot-Jupiter-like exoplanet atmospheres. However, there are some experimental limitations that need to be noted. First, reactions on the wall of the cell can catalyze chemical reactions. Second, we used a static system, where the gases are sealed in the reaction cell, to ensure thermal equilibrium is achieved. However, under such static conditions, the long residence time of the gases in the irradiation area tends to increase the degree of conversion of the reactants to products, resulting in a continuously changing the composition of the gas mixture being photolyzed with time (Clarke et al. 2000). Experiments using a flow system where the reactant gases are continuously renewed would allow us to overcome this issue. In such conditions, the gas mixture composition would reach a steady state after several minutes of irradiation, and a direct quantitative comparisons with planetary atmosphere compositions would be more effective (Peng et al. 2013). However, such studies must ensure thermal equilibrium of gas is reached and need to use highly sensitive spectroscopic techniques (due to lower amounts of product formation) for quantification of spectral signatures.

Finally, the hydrogen lamp that has been used for these experiments does not reproduce total stellar photon flux. Molecules such as $H_2O$ absorb at a wide range of UV wavelengths and photodissociation of water or carbon dioxide might be more efficient in planetary atmospheres than in our experimental setup.

## Conclusion

This work presents a first experimental simulation of chemistry in hot carbon-rich exoplanet atmospheres, where UV photochemistry should play a key role. We have irradiated $H_2$:CO gas mixtures at various temperatures from 300 K to 1500 K with Ly-α (121.6 nm) photons and monitored the evolution of the gas-phase composition using infrared spectroscopy and mass spectrometry.

Thermochemistry leads to the formation of carbon dioxide at each studied temperature, and the formation of methane is observed only occasionally, indicating that either the formation of methane is inefficient, that the formation efficiency has a strong temperature sensitivity, and/or that methane is highly reactive under these conditions, leading to the formation of more complex organics. The observed thermochemistry induces the formation of a new thermal equilibrium composition – different from the initial gas mixture composition used in our experiments – which depends on the gas mixture's temperature.

Further, our results showed that subsequent irradiation of these equilibrium gas mixtures promotes very efficient photochemistry, depleting the gas phase in CO and enhancing the production of carbon dioxide and water. These processes depend on the gas temperatures – at temperatures below ~1200 K, the main photochemical product is $CO_2$, and at higher temperatures, the mixing ratio of $CO_2$ decreases and water becomes the main product. The production of $CO_2$ is explained by the photoexcitation of CO, which reacts with ground state CO to generate C + $CO_2$. This formation of $CO_2$ is also at the origin of the water formation. Indeed, the formation of $O(^1D)$ radicals from $CO_2$ photodissociation leads to the formation of water through subsequent neutral pathways. Carbon produced during the formation of $CO_2$ could react with excess $H_2$ to form $CH_4$. Our initial experimental study shows that photochemistry could efficiently modify the composition of exoplanet atmospheres. In particular, our results highlight that (1) $CO_2$ can be efficiently produced despite competitive destruction processes, in contradiction with current chemical modeling studies (Venot, et al. 2015), and (2) $H_2O$ can be produced via photochemistry in carbon enriched (C/O=1) atmospheres. The second point is important because of the potential for the presence of $H_2O$ in hot Jupiter atmospheres to bias the estimation of the planetary C/O ratio when spectra containing $H_2O$ are interpreted with thermochemical equilibrium models. We expect with decreasing C/O ratio, an increase in both thermal and photochemical transformations to occur, in particular, leading to an increase in production of $H_2O$.

Finally, we have observed the formation of solid organic thin films after the irradiation of the gas phase at 1473 K with Ly-α. We found that non-volatile hydrocarbon aerosols are formed with HCO-functionality involving aromatic and aliphatic hydrocarbons. This result demonstrates that refractory organic aerosols can be formed in hot exoplanet atmospheres with an enhanced C/O ratio. This critical result highlights the importance of conducting laboratory experiments simulating relevant exoplanet environments in order to evaluate the formation and the properties of aerosols in the exoplanet atmospheres. Further studies including other species such as $CO_2$, $CH_4$ and $N_2$ in the initial gas-mixture and with different C/O ratios would improve our understanding of gas and solid aerosol compositions of exoplanet atmospheres.

## Acknowledgments

The research work has been carried out at the Jet Propulsion Laboratory, California Institute of Technology, under a contract with the National Aeronautics and Space Administration. This research work has been supported by the JPL Strategic R&TD funding under "Exoplanet Science Initiative, ESI". We are grateful to Dr. D. Marchione for his contribution to the development of the CAAPSE instrument. The authors would like to thank two anonymous reviewers for their valuable comments which improved the content of this paper.